\documentclass[11pt,showpacs,preprintnumbers,amsmath,aps,amssymb]{revtex4}
\usepackage{graphicx}
\usepackage{dcolumn}
\usepackage[english]{babel}
\usepackage{bm}
\usepackage[font=scriptsize]{caption}
\usepackage{ragged2e}
\usepackage[font=small,labelfont=bf,justification=raggedright]{caption}
\usepackage[utf8]{inputenc}
\usepackage[tikz]{bclogo}
\usepackage[framemethod=tikz]{mdframed}
\usepackage{mathtools}
\usepackage{amsmath}     
\usepackage{array}       
\usepackage{arydshln}   
\usepackage{xcolor}
\usepackage{lscape}
\usepackage{cancel}
\usepackage{rotating}
\colorlet{BLUE}{blue} \colorlet{RED}{red}
\newsavebox{\measurebox}

\begin{document}
\title{Hamiltonian formalism for a 5-D brane cosmolgy}

\author{J. Socorro$^1$ }
\email{socorro@fisica.ugto.mx}
\author{Juan Luis P\'erez$^1$ }
\email{jl.perezp@ugto.mx}

\affiliation{ $^1$ Department of Physics,  Division of Science and
Engineering, University of Guanajuato, Campus Le\'{o}n,
     Le\'{o}n 37150,  Mexico\\
}
\begin{abstract}
In this work we explores the dynamic solutions of a brane system in a five dimensional space-time under the Hamiltonian perspective, 
using the De Donner Weyl formalism. We obtain the dynamic equations for the scale factors, and obtain the Friedmann-like equations in general way. 
We present the solutions of the scale factors as a function of the extra coordinate, $y$, we show that the geometry is related to the matter-energy component in the brane. Finally,
 we develop the ADM formalism in the brane point of view with the foliation of the 3-dimensional space which is embedded in a 5-dimensional manifold. \\

\end{abstract}

\maketitle

\section{Introduction}

The possibility that the gravitational interaction propagates in a multi-dimensional space-time, constitutes one of the more relevant scenarios for exploring extensions of General Relativity. 
In brane world models, the observable universe is considered as a four-dimensional hypersurface (brane), embedded in a space-time of more than 4 dimensions (bulk). 
Whereas non gravitational fields (matter and interactions) can be confined to the brane, gravity must propagate through the additional dimensions, resulting in an effective modified gravitational dynamic \cite{MaartensKoyama2010}. 
A particularly important realization corresponds to Randall-Sundrum models, in which, a brane is founded embedding in 5-D space-time with a negative cosmological constant \cite{RandallSundrum1999a,RandallSundrum1999b}.
 Projection of the 5-D Einstein equations onto the brane leads to effective gravitational theory that contains quadratic terms in the energy-momentum tensor and other contributions associated with the Weyl tensor in 5 dimensions \cite{ShiromizuMaedaSasaki2000,Maeda2002}. In particular, the projection of the Weyl tensor contains information about the geometry of the bulk, that cannot be determined from the intrinsic equations in the brane \cite{MaartensKoyama2010}.
A natural tool to study the gravitational dynamic of a foliated space-time is the Arnowit-Deser-Misner (ADM) formalism \cite{ArnowittDeserMisner1962}. In this formalism, the whole space-time is decomposed into a family of hypersurfaces, and the metric is expressed in terms of the induced metric, lapse function and the shift vector. Extrinsic curvature allows us to separate both: gravitational evolution equations and restrictions. Although originally the model was formulated by means of a temporal foliation, the geometrical nature of this construction permits the use of a spatial coordinate as a foliation parameter. Extrinsic curvature then acquires a central role in the description of the geometry of the brane and in the implementation of the Israel junction conditions \cite{Israel1966}.
From a different perspective, the De Donner-Weyl formalism provides a Hamiltonian-type covariant formulation in field theories, in which the space-time coordinates are treated in a symmetrical manner and a temporal privileged direction is not required. In this approach, canonical momenta are generalized (elevated) to a polycanonical momenta, associated with derivative of the field with respect to each spatial-time coordinate. This structure is of particular interest in gravitational theories  and in covariant formulations of fields dynamics, and offers a natural point of comparison with ADM formulation.
The relation between these two formulations is especially relevant in braneworld scenarios. While the ADM formalism allows the interpretation of the additional dimension as a direction of geometric evolution and to formulate the problem in terms of induced metric and extrinsic curvature, the De Donner-Weyl formalism provides a covariant perspective about the Hamiltonian structure of the theory. Comparison between both approaches can contribute to clarifying what part of the gravitational dynamic is associated with the elected foliation and what features are independent of such election.
In this work we develop an ADM formulation of the gravitational dynamic of 5-D space-time in the context of brane worlds, taking both additional dimension and time as directions associated with the foliation. Starting from the Einstein-Hilbert action, we study the structure of the induced metric, lapse functions, extrinsic curvature and the equations of restrictions. The goal of this work is to establish a formulation that allows us to relate the explicit dynamic in the bulk with the induced geometry in the context of the covariant description of the De Donner-Weyl formalism \cite{DeDonder1930,Weyl1935}.
The paper is organized in the following way: In section II we explore the basic mathematical tools that are necessary for our study: the notation and initial hypothesis and the solutions of Einstein equations for a generalized Randall-Sundrum model. Section III is devoted to analyzing the Hamiltonian formalism for the model, and with the aid of the De Donner-Weyl formalism, we find a solution for the metric coefficients as a function of y. In Section IV we develop the ADM formalism in which the foliation is carried out in a direction that can be expressed as a linear combination of time and extra dimension. Perspectives and conclusions are presented in the final section of the article.

\section{Mathematical background}

\subsection{Covariant Hamiltonian density}

Consider a Lagrangian that depends on $m\times n$ fields $\phi_{ij}$ that are functions of $M+1$ space-time variables $x^\mu=\{x^0,x^1,x^2,...,x^M\}$. 
The Lagrangian depends on the fields $\phi_{ij}$, their derivatives $\partial_\mu \phi_{ij} \equiv \partial \phi_{ij} / \partial x^\mu$, and the variables $x^\mu$.

\begin{equation}\label{lagrangiano}
L=\int \mathcal{L} (\phi_{ij}, \partial_\mu \phi_{ij}, x^\mu), 
\end{equation}
from this Lagrangian, we define the conjugate momenta

\begin{equation}
\Pi^{(\mu)ij} \equiv \frac{\partial \mathcal{L}}{\partial (\partial_\mu \phi_{ij})},
\end{equation}
in this work, the index associated with the coordinate will be written in parentheses. The corresponding Hamiltonian density is written as
\begin{equation}
\mathcal{H}(\phi_{ij}, \Pi^{(\mu)ij}, x^\mu)=\Pi^{(\mu)ij} \partial_\mu \phi_{ij}- \mathcal{L} (\phi_{ij}, \partial_\mu \phi_{ij}, x^\mu). 
\end{equation}

The Hamilton's equations are, in this case,

\begin{eqnarray}
\frac{\partial \mathcal{H}}{\partial \Pi^{(\mu)ij}} = \frac{\partial \phi_{ij}}{\partial x^\mu}, \ \ \  \ \ \ 
\frac{\partial \mathcal{H}}{\partial \phi_{ij}} = -\frac{\partial \Pi^{(\alpha)ij}}{\partial x^\alpha},
\end{eqnarray}
in the previous equations we use the repeated-index convention. This is known in the literature as the De Donder-Weyl formalism \cite{StruckmeierRedelbach2008}, 
and it will be useful in Section III and IV, when we obtain the dynamical equations on the brane.

\subsection{A brief introduction to the ADM formalism}
\label{sec}

For the ADM formalism \cite{ArnowittDeserMisner1962}, consider a $(d+1)$-dimensional space-time $\mathcal{M}$ foliated by a family of $d$-dimensional hypersurfaces $\Sigma_t$. The metric can be written in the form
\begin{equation}
	ds^2=-N^2dt^2+h_{ij}\left(dx^i+N^i dt\right)\left(dx^j+N^j dt\right),
\end{equation}
where $h_{ij}$ is the induced metric on the hypersurfaces $\Sigma_t$, $N$ is the \textit{lapse} function, and $N^i$ is the \textit{shift} vector.
The unit normal vector to the hypersurfaces is given by
\begin{equation}
	n_\mu=(-N,0,\ldots,0),
	\qquad
	n^\mu=\left(\frac{1}{N},-\frac{N^i}{N}\right),
	\qquad
	n^\mu n_\mu=-1.
\end{equation}

The way in which each hypersurface is embedded in space-time is characterized by its extrinsic curvature,
\begin{equation}
	K_{ij}=\frac{1}{2N}\left(\dot{h}_{ij}-D_iN_j-D_jN_i\right),
\end{equation}
where the dot denotes a derivative with respect to $t$ and $D_i$ is the covariant derivative compatible with the induced metric $h_{ij}$. The trace of the extrinsic curvature is defined by $K=h^{ij}K_{ij}.$

The geometric decomposition of the $(d+1)$-dimensional Ricci scalar in terms of quantities intrinsic and extrinsic to the hypersurfaces leads to
\begin{equation}
	{}^{(d+1)}R={}^{(d)}R+K_{ij}K^{ij}-K^2+\text{boundary terms},
\end{equation}
where ${}^{(d)}R$ is the Ricci scalar constructed from the induced metric $h_{ij}$. Consequently, the Einstein--Hilbert action,
\begin{equation}
	S_{EH}=\frac{1}{2\kappa_{d+1}^2}\int_{\mathcal{M}}d^{d+1}x\sqrt{-g}\left({}^{(d+1)}R-2\Lambda_{d+1}\right),
\end{equation}
can be expressed, after appropriately separating the boundary terms, in the ADM form
\begin{equation}
	S_{EH}=\frac{1}{2\kappa_{d+1}^2}\int dt \ d^d x,N\sqrt{h}\left[{}^{(d)}R+K_{ij}K^{ij}-K^2-2\Lambda_{d+1}\right].
\end{equation}

This expression is particularly useful because the time derivatives of the induced metric appear only through the extrinsic curvature. In this way, the gravitational action can be formulated in terms of variables that allow its canonical structure to be identified directly.

The momentum canonically conjugate to $h_{ij}$ is given by
\begin{equation}
	\pi^{ij}=\frac{\partial\mathcal{L}_{ADM}}{\partial\dot{h}_{ij}}=\frac{\sqrt{h}}{2\kappa_{d+1}^2}\left(K^{ij}-Kh^{ij}\right).
\end{equation}

The variation of the action with respect to the \textit{lapse} function and the \textit{shift} vector leads, respectively, to the Hamiltonian constraint and the momentum constraint. In the absence of matter, these equations can be written as
\begin{eqnarray}
	{}^{(d)}R+K^2-K_{ij}K^{ij}-2\Lambda_{d+1}&=&0, \\
	D_j\left(K^{ij}-Kh^{ij}\right)&=&0.
\end{eqnarray}

The remaining equations determine the evolution of the induced metric and the extrinsic curvature from one hypersurface to the next. Thus, the ADM formalism separates the gravitational equations into a set of constraints that must be satisfied on each hypersurface and a set of equations governing their evolution.

\subsection{Friedmann equations}
Hamilton's equations in the ADM formalism are equivalent to Einstein's equations, since all dynamical information is contained in the canonical variables and their momenta, then in this part of the work, we present the usual calculus for solve the Einstein equation in four dimensions. The line element for the Friedmann.Robertson-Walker become 

\begin{equation}
	ds^2_4=-N^2(t)dt^2+A^2(t)\left[\frac{dr^2}{1-\kappa r^2}+r^2 d\theta^2 + r^2 sin^2 \theta d\phi^2\right],
\end{equation}
obtaining the corresponding scalar curvature  
\begin{equation}\label{curvFRW}
R=\frac{6}{N^2}\left[ \frac{\ddot{A}}{A}+\frac{\dot{A}^2}{A^2}-\frac{\dot{A}}{A}\frac{\dot{N}}{N}+\kappa\frac{N^2}{A^2} \right],
\end{equation}
thus, we write the Einstein-Hilbert Lagrangian with a cosmological constant and a matter contribution in general relativity:
\begin{equation}
L=\int{d^3x\sqrt{-g}\left[R-2\Lambda+\mathcal{L}_m \right]}=\int{dVA^3N\left[R-2\Lambda+\mathcal{L}_m \right]},
\end{equation}
where $dV=d^3x(r^2sin\theta/\sqrt{1-\kappa r^2})$. Therefore, $A^3N\left[R+\mathcal{L}_m \right]$ is the Lagrangian density:
\begin{equation}\label{denslagFRW}
	\mathcal{L}=A^3N\left[R-2\Lambda+\mathcal{L}_m \right]
\end{equation}
$\mathcal{L}_m$ is the Lagrangian corresponding to the energy distribution of a perfect fluid, $\mathcal{L}_m=-16\pi G\rho$, where $\rho$ depends on the scale factor using a energy-momentum tensor of barotropic perfect fluid,
\begin{equation}\label{densityFRW}
\rho=M_{\gamma}A^{-3(1+\gamma)}.
\end{equation}

Substituting (\ref{curvFRW}) and (\ref{densityFRW}) into (\ref{denslagFRW}) and eliminating terms with second derivatives that do not contribute to the action,
\begin{equation}
	\mathcal{L}=-6\frac{A(\dot{A})^2}{N} -16\pi G M_{\gamma}A^{-3\gamma}N-2A^3N\Lambda+6\kappa AN.
\end{equation}

Since the Lagrangian depends only on $A$ and $\dot{A}$, we define the canonical momentum associated with the variable $A$ as
\begin{equation}
	\Pi_A=\frac{\partial \mathcal{L}}{\partial \dot{A}}=-12\frac{A \dot{A}}{N}.
\end{equation}
We solve for $\dot{A}$ and substitute it into the expression for the relativistic Hamiltonian,
\begin{eqnarray}\nonumber
	\mathcal{H}&=&\dot{A}\Pi_A-\mathcal{L}\\
                           &=&	-\frac{N}{24A^3}\left[ A^2\Pi^2_A-384\pi G M_{\gamma}A^{3(1-\gamma)}+144A^4\kappa-48A^6\Lambda \right].
\end{eqnarray}

Now let us use the notation $A=e^{a}$ and $P_a=A\Pi_A$, as in \cite{SocorroPerez2026}.
\begin{equation}
	\mathcal{H}=-\frac{1}{24}e^{-3a}\left[ P^2_a-384\pi G M_{\gamma}e^{3a(1-\gamma)}+144e^{4a}\kappa-48e^{6a}\Lambda \right],
\end{equation}
Hamilton's equations are invariant under this transformation, so
\begin{eqnarray}
	\dot{a}&=&\frac{\partial \mathcal{H}}{\partial P_a}=-\frac{1}{12}e^{-3a}P_a \label{apuntoFRW}\\
	\dot{P_a}&=&-\frac{\partial \mathcal{H}}{\partial a}=e^{-3a}\left[-48\pi G(1-\gamma)M_{\gamma}e^{3a(1-\gamma)}+24e^{4a}\kappa-12e^{6a}\Lambda\right] \label{PapuntoFRW}
\end{eqnarray}
where we have used the fact that in Relativity, $N\mathcal{H}=0$. We can clearly identify the first equation as the Friedmann equation, which is obtained by squaring and using the Hamiltonian constraint
\begin{eqnarray}\nonumber
	\dot{a}^2&=&\frac{1}{144}e^{-6a}P^2_a\\ \nonumber
	&=&\frac{1}{144}e^{-6a}\left[384\pi G M_{\gamma}e^{3a(1-\gamma)}-144e^{4a}\kappa+48e^{3a}\Lambda \right]\\
	&=&\frac{8}{3}\pi G M_{\gamma}e^{-3(1+\gamma)}-\kappa e^{-2a}+\frac{1}{3}\Lambda.
\end{eqnarray}

Now, returning to the original variables, $A=e^{a}$ and using the definition of $\rho$
\begin{equation}
	\frac{\dot{A}^2}{A^2}=H^2=\frac{8}{3}\pi G\rho-\frac{\kappa}{A^2}+\frac{1}{3}\Lambda,
\end{equation}
we can also obtain the acceleration equation by differentiating (\ref{apuntoFRW}) and combining it with (\ref{PapuntoFRW})
\begin{equation}
\frac{\ddot{A}}{A}-\frac{\dot{A}^2}{A^2}=-4\pi G(1+\gamma)\rho+\frac{\kappa}{A^2}.
\end{equation}
The virtue of the Hamiltonian method is that it provides a system of first-order equations, in contrast to Einstein's equations. In \cite{SocorroPerez2026,SocorroGilPerez2026} it has already been used to explore dynamical solutions in $F(R)$ theories and with potential terms, and in this work we will analyze the system of solutions for a 5-dimensional model.

\subsection{5-D Field equations}

In previous works \cite{PerezUrenaCordero2015,PerezGarciaUrena2014}, a two-brane system has been studied, and the correlation between the matter-energy components on each brane is analyzed; in this work, we will conduct a similar analysis, but in the context of single-brane models.. The most general line element for Randall-Sundrum-type models can be considered in the form
\begin{equation}\label{metrica5d}
ds^2_5=-N^2 (t,y)dt^2+A^2 (t,y)\gamma_{ij} dx^i dx^j +B^2 (t,y) dy^2 
\end{equation}

In this case, we can consider the matter fields to be confined to a 4-dimensional hypersurface called the brane, which is fixed at $y=0$, and the additional coordinate, $y$, is compactified and has a reflection symmetry,
\begin{eqnarray}
  \label{tme} T_{AB}= -\frac{\Lambda_5}{k^{2}_{5}} g_{AB}+
  \frac{\delta(y)}{B_{0}} {\rm diag}(-\rho_{0},\textbf{p}_{0},0).
\end{eqnarray}

Under these assumptions, and assuming that the matter-energy in the brane is a perfect fluid, the following boundary conditions are satisfied \cite{WangCaiSantos2008}:
\begin{eqnarray}
&&\frac{2}{B_0}\left[ 2\frac{\left.A^{\prime}\right|_0}{A_0}+\frac{\left.N^{\prime}\right|_0}{N_0} \right]=+k_5 (p_0-\lambda_0)\\ \label{junction-an}
&&\frac{6}{B_0}\left[ \frac{\left.A^{\prime}\right|_0}{A_0} \right]=-k_5 (\rho_0+\lambda_0) \label{junction-a}
\end{eqnarray}
where $\left.F^{\prime}\right|_0$ denotes the derivative with respect to the variable $y$, evaluated in the vicinity of $y=0+$.

The five-dimensional non zero Einstein tensors, $G_{_{AB}}$, for the metric (\ref{metrica5d}) are written as a non-linear differential equations system
\begin{eqnarray}
\label{einstein0}
\tilde{G}_{00}&=&3\frac{\dot{A}}{A}\left(\frac{\dot{A}}{A}+\frac{\dot{B}}{B}\right)-3\frac{N^{2}}{B^{2}}\left[\frac{A''}{A}+\frac{A'}{A}\left(\frac{A'}{A}-\frac{B'}{B}\right)\right]+3\kappa \frac{N^2}{A^2}, \\ 
\label{einstein1}
\tilde{G}_{ij}&=&\frac{A^{2}}{B^{2}}\gamma_{ij}\left\{\frac{A'}{A}\left(\frac{A'}{A}+2\frac{N'}{N}\right)-\frac{B'}{B}\left(\frac{N'}{N}+2\frac{A'}{A}\right)\right\}\\
&+&\frac{A^{2}}{B^{2}}\gamma_{ij}\left\{2\frac{A''}{A}+\frac{N''}{N}\right\}+\frac{A^{2}}{N^{2}}\gamma_{ij}\left\{\frac{\dot{A}}{A}\left(-\frac{\dot{A}}{A}+2\frac{\dot{N}}{N}\right)\right\}\nonumber\\
&+&\frac{A^{2}}{N^{2}}\gamma_{ij}\left\{-2\frac{\ddot{A}}{A}+\frac{\dot{B}}{B}\left(-2\frac{\dot{A}}{A}+\frac{\dot{N}}{A}\right)-\frac{\ddot{B}}{B}\right\}-\kappa \gamma_{ij}, \\
\label{einstein2}
\tilde{G}_{04}&=&3\left(\frac{\dot{A}}{A}\frac{N'}{N}+\frac{\dot{B}}{B}\frac{A'}{A}-\frac{\dot{A}'}{A}\right), \\
\label{einstein3}
\tilde{G}_{44}&=&3\frac{A'}{A}\left(\frac{A'}{A}+\frac{N'}{N}\right)-3\frac{B^{2}}{N^{2}}\left[\frac{\ddot{A}}{A}+\frac{\dot{A}}{A}\left(\frac{\dot{A}}{A}-\frac{\dot{N}}{N}\right)\right]-3\kappa \frac{B^2}{A^2}.
\end{eqnarray}

Before to continue, it is necessary to rewrite the above equations in a more convenient way by using the new variables,
\begin{equation}\label{xyzw}
X=\frac{A\dot{A}}{N}, \ \ Y=\frac{AA^{\prime}}{B}, \ \ Z=\frac{(AN)^{\prime}}{B}, \ \ W=\frac{\dot{(AB)}}{N}.
\end{equation}
In the interval $0<y<y_{c}$, where Einstein's equatios are valid, $G_{AB}=\kappa^{2}_{(5)}T_{AB}$, reads
\begin{eqnarray}\label{einstein-compactas}
\label{e-c-0}
\frac{Y'}{B}-\frac{XW}{AB}&=&k-\frac{1}{3}A^{2}\Lambda_{5},\\
\label{e-c-1}
\frac{Y'}{B}-\frac{\dot{X}}{N}+\frac{A}{BN}(Z'-\dot{W})&=&k-A^{2}\Lambda_{5},\\
\label{e-c-2}
W&=&A\frac{X'}{Y},\\ 
\label{e-c-3}
Z&=&A\frac{\dot{Y}}{X},\\
\label{e-c-4}
\frac{\dot{X}}{N}-\frac{YZ}{AN}&=&-k+\frac{1}{3}A^{2}\Lambda_{5},
\end{eqnarray}
to find the modified Friedmann equation, substituting (\ref{e-c-2}) in (\ref{e-c-0}) and integrating, we obtain
\begin{equation}\label{fr-3}
Y^{2}-X^{2}=A^{2}\kappa-A^{4}\frac{\Lambda_{5}}{6}+\emph{constant},
\end{equation}
now, substituting the values of $X$ and $Y$, we obtain the equation
\begin{eqnarray}
	\left[\frac{A A^{\prime}}{B}\right]^2 -\left[\frac{A \dot{A}}{N}\right]^2-\kappa A^2+\frac{1}{6}\Lambda_5 A^4&=&constant.
\end{eqnarray}

This equation holds throughout the 5-dimensional space, in particular on the brane. Evaluating at $y=0$ and taking into account the junction conditions (\ref{junction-a}),
\begin{eqnarray}
	H^2_0=k^2_4\rho_0\left[1+\frac{\rho_0}{2\lambda_0}\right] +\frac{\Lambda_4}{3}-\frac{\kappa}{A^2}+\frac{cte}{A^4}
\end{eqnarray}
where $k^2_4=\frac{\lambda_0 k^2_5}{18}$, $\Lambda_4=\frac{k^2_5\lambda^2_0}{12}+\frac{\Lambda_5}{2}$. Now, if we add equations (\ref{e-c-0}) and (\ref{e-c-4}), multiply by $ABN$, and rewrite $ANY^{\prime}=(ANY)^{\prime}-(AN)^{\prime}Y$ and $AB\dot{X}=(ABX)^{\centerdot}-X(AB)^{\centerdot}$, we obtain,
\begin{equation}\label{identidad}
	3(ANY)^{\prime}-3(ABX)^{\centerdot}-6ABN\kappa +2A^3BN\Lambda_5=0.
\end{equation}

In the following section we will see that this expression is the Hamiltonian constraint $\mathcal{H}$, where $ANY$ and $ABX$ are the canonical momenta of the theory.

\section{Hamiltonian formalism for 5D cosmology}

In a 5-D model with metric (\ref{metrica5d}), the Einstein-Hilbert action is written as
\begin{eqnarray} \nonumber
S&=&\int{d^5x\sqrt{-g}(R-2\Lambda_5)}\\
&=&\int{dt\int{d^4V A^3BN(R-2\Lambda_5)}} 
\end{eqnarray}
The Lagrangian density is then defined as $\mathcal{L}=A^3BN(R-2\Lambda_5)$. Varying the action leads to Einstein's equations $G_{AB}=\Lambda_5 g_{AB}$. All our variables are 5-dimensional unless otherwise indicated.

%

First, let us calculate the corresponding curvature scalar, $R$
\begin{eqnarray}\nonumber
	R&=&\frac{6\kappa}{A^2}+\frac{6}{N^2}\left[ \frac{\ddot{A}}{A}+\left(\frac{\dot{A}}{A}\right)^2+\frac{\dot{A}}{A}\frac{\dot{B}}{B}-\frac{\dot{A}}{A}\frac{\dot{N}}{N}+\frac{1}{3}\frac{\ddot{B}}{B}-\frac{1}{3}\frac{\dot{B}}{B}\frac{\dot{N}}{N}\right]\\
	&&-\frac{6}{B^2}\left[ \frac{A^{\prime\prime}}{A}+\left(\frac{A^{\prime}}{A}\right)^2+\frac{A^{\prime}}{A}\frac{N^{\prime}}{N}-\frac{A^{\prime}}{A}\frac{B^{\prime}}{B}+\frac{1}{3}\frac{N^{\prime\prime}}{N}-\frac{1}{3}\frac{B^{\prime}}{B}\frac{N^{\prime}}{N}\right],
\end{eqnarray}
Here a dot denotes a derivative with respect to time $t$ and a prime denotes a derivative with respect to the coordinate $y$. The term corresponding to the curvature scalar in the Lagrangian is $(A^3BN)R$, and once terms with second derivatives, that do not contribute to the action, have been eliminated, it becomes:
\begin{equation}
	(A^3BN)R=-\frac{6A^3B}{N}\left[ \left( \frac{\dot{A}}{A}\right)^2+\frac{\dot{A}}{A}\frac{\dot{B}}{B}\right]  
	+\frac{6A^3N}{B}\left[ \left( \frac{{A^{\prime}}}{A}\right)^2+\frac{{A^{\prime}}}{A}\frac{{N^{\prime}}}{N}\right] +6\kappa ABN
\end{equation}

Thus, the Lagrangian density is written as
\begin{equation}
	\mathcal{L}=-\frac{6A^3B}{N}\left[ \left( \frac{\dot{A}}{A}\right)^2+\frac{\dot{A}}{A}\frac{\dot{B}}{B}\right]  
	+\frac{6A^3N}{B}\left[ \left( \frac{{A^{\prime}}}{A}\right)^2+\frac{{A^{\prime}}}{A}\frac{{N^{\prime}}}{N}\right] +6\kappa ABN
	-2\Lambda_5 A^3BN 
\end{equation}
Note that $\mathcal{L}=\mathcal{L}(A, B, N, \dot{A}, \dot{B}, A^{\prime}, N^{\prime})$, so we define the generalized momentum variables
\begin{equation}
	\Pi^0_A=\frac{\partial \mathcal{L}}{\partial \dot{A}}\ , \ \
	\Pi^4_A=\frac{\partial \mathcal{L}}{\partial A^{\prime}}\ , \ \
	\Pi^0_B=\frac{\partial \mathcal{L}}{\partial \dot{B}}\ , \ \
	\Pi^4_N=\frac{\partial \mathcal{L}}{\partial N^{\prime}} \ \
\end{equation}

The superscript 0 refers to the coordinate $t$, while the index 4 refers to the coordinate $y$. Using the expression for the Lagrangian density, the canonical momenta are written as
\begin{eqnarray}
	\Pi^0_A&=&-\frac{6A^3B}{N}\left[ 2\frac{\dot{A}}{A^2}+\frac{\dot{B}}{AB} \right],\\
	\Pi^4_A&=&\frac{6A^3N}{B}\left[ 2\frac{A^{\prime}}{A^2}+\frac{N^{\prime}}{AN} \right],\\
	\Pi^0_B&=&-\frac{6A^3B}{N}\left[ \frac{\dot{A}}{AB} \right], \\
	\Pi^4_N&=&\frac{6A^3N}{B}\left[ \frac{A^{\prime}}{AN} \right]. 
\end{eqnarray}

From these equations, we can solve for the derivatives in order to write the Hamiltonian solely in terms of the momenta and variables. As seen in the previous section, the Hamiltonian in the De Donner-weyl formalism is written as
\begin{eqnarray}\nonumber
	\mathcal{H}&=&\dot{A}\Pi^0_A+A^{\prime}\Pi^4_A+\dot{B}\Pi^0_B+N^{\prime}\Pi^4_N-\mathcal{L}\\ \nonumber
	\mathcal{H}&=&\frac{N}{6A^3B}\left[ (B\Pi^0_B)^2-(A\Pi^0_A)(B\Pi^0_B) \right]-\frac{B}{6A^3N}\left[ (N\Pi^4_N)^2-(A\Pi^4_A)(N\Pi^4_N) \right]\\
	&&-6\kappa ABN+2\Lambda A^3BN
\end{eqnarray}
For convenience, and to simplify some calculations, we can rewrite this expression in terms of the variables $Q=e^q$, $Q\Pi^{\mu}_Q=P^{\mu}_q$, $Q=\{A, B, N\}$.
\begin{eqnarray} \nonumber
	\mathcal{H}&=&\frac{1}{6}e^{-(3a+b)+n}\left[(P^0_b)^2- P^0_a P^0_b \right]-\frac{1}{6}e^{-(3a+n)+b}\left[ (P^4_n)^2-(P^4_a)(P^4_n) \right]\\
	&&-6\kappa e^{a+b+n}+2\Lambda e^{3a+b+n}
\end{eqnarray}

In these new variables, Hamilton's equations are invariant, so the Donder-Weyl equations hold,
\begin{eqnarray}
	&&a^{\prime}=\frac{\partial \mathcal{H}}{\partial P^4_a}, \ \ \ \dot{a}=\frac{\partial \mathcal{H}}{\partial P^0_a}, \ \ \  n^{\prime}=\frac{\partial \mathcal{H}}{\partial P^4_n}, \ \ \ \dot{b}=\frac{\partial \mathcal{H}}{\partial P^0_b},\\ \nonumber
	\\
	&&(P^0_a)^{\centerdot}+(P^4_a)^{\prime}=-\frac{\partial \mathcal{H}}{\partial a}, \ \ \ (P^0_b)^{\centerdot}=-\frac{\partial \mathcal{H}}{\partial b}, \ \ \ (P^4_n)^{\prime}=-\frac{\partial \mathcal{H}}{\partial n}
\end{eqnarray}

Using the Hamiltonian found above, and assuming that $\mathcal{H}\neq 0$,

\begin{eqnarray}
	&&a^{\prime}=\frac{1}{6}e^{-(3a+n)+b}P^4_n \label{a-prima}\\
	&&n^{\prime}=-\frac{1}{6}e^{-(3a+n)+b}\left[2P^4_n-P^4_a\right] \label{n-prima}\\
	&&\dot{a}=-\frac{1}{6}e^{-(3a+b)+n}P^0_b \label{a-punto} \\
	&&\dot{b}=\frac{1}{6}e^{-(3a+b)+n}\left[2P^0_b-P^0_a \right] \label{b-punto} \\
	&&(P^4_n)^{\prime}=-\frac{1}{3}e^{-(3a+n)+b}\left[(P^4_n)^2-P^4_a P^4_n\right]-\mathcal{H} \label{pn4-prima} \\
	&&(P^0_b)^{\centerdot}=\frac{1}{3}e^{-(3a+b)+n}\left[(P^0_b )^2-P^0_a P^0_b\right]-\mathcal{H} \label{pb0-prima}\\
	&&(P^0_a)^{\centerdot}+(P^4_a)^{\prime}=-24\kappa e^{a+b+n}+12\Lambda e^{3a+b+n}-3\mathcal{H}
\end{eqnarray}

These equations are equivalent (and fully consistent) with Einstein's equations (\ref{einstein0}) to (\ref{einstein3}); the last three equations correspond to $G_{44}$, $G_{00}$, and $G_{ij}$, respectively. However, the equation $G_{04}$ is not present and there is no way to obtain it, so in this theory it is not possible to obtain the Friedmann equation directly. Despite this inconsistency between the two formalisms, we can obtain the Friedmann equation artificially if we impose that the Hamiltonian vanishes. For this purpose, we can rewrite the Hamiltonian in terms of $(P^4_n)^{\prime}$ and $(P^0_b)^{\centerdot}$
\begin{eqnarray}
\mathcal{H}&=&\frac{1}{2}(P^4_n)^{\prime}+\frac{\mathcal{H}}{2}+\frac{1}{2}(P^0_b)^{\centerdot}+\frac{\mathcal{H}}{2}-6\kappa e^{a+b+n}+2\Lambda e^{3a+b+n}\nonumber \\ \label{hamconst}
	0&=&\frac{1}{2}(P^4_n)^{\prime}+\frac{1}{2}(P^0_b)^{\centerdot}-6\kappa e^{a+b+n}+2\Lambda e^{3a+b+n}
\end{eqnarray}

This last expression is precisely the Hamiltonian constraint, which is equivalent to equation (\ref{identidad}) and can also be derived by taking the functional derivatives $\delta S / \delta N =0$ and $\delta S / \delta B =0$. 

If we combine (\ref{pn4-prima}) with (\ref{a-prima}) and (\ref{n-prima}) we obtain $(P^4_n)^{\prime}=2P^4_n(a^{\prime}+n^{\prime})-\mathcal{H}$; similarly, if we combine (\ref{pb0-prima}) with (\ref{a-punto}) and (\ref{b-punto}), $(P^0_b)^{\centerdot}=2P^0_b(\dot{a}+\dot{b})-\mathcal{H}$. Defining the variables $Y\equiv e^{-(a+n)}P^4_n/6 $ and $X\equiv e^{-(a+b)}P^0_b/6 $, and differentiating with respect to $y$ and with respect to $t$, we obtain
\begin{eqnarray} \nonumber
6Y^{\prime}&=&e^{-(a+n)}(P^4_n)^{\prime}-e^{-(a+n)}P^4_n(a^{\prime}+n^{\prime}) \\ \nonumber
&=&e^{-(a+n)}(P^4_n)^{\prime}-\frac{1}{2}e^{-(a+n)}(P^4_n)^{\prime}-\frac{1}{2}e^{-(a+n)}\mathcal{H} \\
&=&\frac{1}{2}e^{-(a+n)}(P^4_n)^{\prime}-\frac{1}{2}e^{-(a+n)}\mathcal{H} \\ \nonumber
6\dot{X}&=&e^{-(a+b)}(P^0_b)^{\centerdot}-e^{-(a+b)}P^0_b(\dot{a}+\dot{b}) \\ \nonumber
&=&e^{-(a+b)}(P^0_b)^{\centerdot}-\frac{1}{2}e^{-(a+b)}(P^0_b)^{\centerdot} -\frac{1}{2}e^{-(a+b)}\mathcal{H}\\ 
&=&\frac{1}{2}e^{-(a+b)}(P^0_b)^{\centerdot}-\frac{1}{2}e^{-(a+b)}\mathcal{H}
\end{eqnarray}

Now, let us substitute this result into the Hamiltonian constraint (\ref{hamconst}) and maintaining $\mathcal{H}\neq0$
\begin{eqnarray}
	0&=&\mathcal{H}+6e^{a+n}Y^{\prime}+6e^{a+b}\dot{X}-6\kappa e^{a+b+n}+2\Lambda e^{3a+b+n}
\end{eqnarray}

Let us use the chain rule: $Y^{\prime}=a^{\prime}Y_a$ and $\dot{X}=\dot{a}X_a$, and using equations (\ref{a-prima}) and (\ref{a-punto}), as well as the definitions of $X$ and $Y$, we finally obtain
\begin{eqnarray}
	0&=&\mathcal{H}e^{-(b+n)+a}+2Y Y_a -2 X X_a-2\kappa e^{2a}+\frac{2}{3}\Lambda e^{4a}
\end{eqnarray}

This equation is perfectly integrable only if we impose $\mathcal{H}=0$, and its solution is
\begin{eqnarray}
	Y^2 -X^2-\kappa e^{2a}+\frac{1}{6}\Lambda e^{4a}&=&constant
\end{eqnarray}

If we return to the notation $A=e^a$ and $B=e^b$, with the variables $Y=A A^{\prime}/B$ and $X=A \dot{A}/N$, we finally obtain
\begin{eqnarray}
	\left[\frac{A A^{\prime}}{B}\right]^2 -\left[\frac{A \dot{A}}{N}\right]^2-\kappa A^2+\frac{1}{6}\Lambda A^4&=&constant.
\end{eqnarray}

This shows that the Hamiltonian formalism is consistent with Einstein's equations. As mentioned above, in general the Hamiltonian $\mathcal{H}$ does not vanish on the surface, and the constraint $\mathcal{H}=0$ is locally compatible with $G_{04}=0$, although in this model such equivalence cannot be demonstrated.
%
%

\subsection{Global solutions in the bulk}

In the analysis presented in \cite{WangCaiSantos2008}, matter is contained on a hypersurface called the brane, such that the energy density has a Dirac delta distribution, so that the derivative of the scale factors $A$ and $N$ with respect to the coordinate $y$ satisfies:
\begin{eqnarray}
&&\frac{2}{B_0}\left[ 2\frac{\left.A^{\prime}\right|_0}{A_0}+\frac{\left.N^{\prime}\right|_0}{N_0} \right]=+k_5 p_0\\
&&\frac{6}{B_0}\left[ \frac{\left.A^{\prime}\right|_0}{A_0} \right]=-k_5 \rho_0
\end{eqnarray}

where the subscript 0 indicates the variable evaluated at $y=0$. Rewriting in the new variables $a, b, n$,
\begin{eqnarray}
&&2e^{-b_0}\left( 2\left.a^{\prime}\right|_0+\left.n^{\prime}\right|_0\right)=-k_5 p_0\\
&&6e^{-b_0}\left( \left.a^{\prime}\right|_0 \right)=+k_5 \rho_0
\end{eqnarray}

Using equations (\ref{a-prima}) and (\ref{n-prima}),
\begin{eqnarray}
\left.P^4_n\right|_0&=&-k_5e^{3a_0+n_0}\rho_0\\
\left.P^4_a\right|_0&=&+3k_5e^{3a_0+n_0}p_0
\end{eqnarray}

From this we can deduce that if the matter field on the brane is a perfect fluid, $p_0=\gamma_0\rho_0$, then the momenta corresponding to the variables $a$ and $n$ are related by
\begin{equation}
\left.P^4_a\right|_0 = -3\gamma_0 \left.P^4_n\right|_0
\end{equation}

In general, we will assume as an Ansatz that this relation holds throughout the 5-dimensional space. Therefore,
\begin{equation}\label{Pa4}
P^4_a= -3\gamma_0 P^4_n
\end{equation}

We can now integrate equation (\ref{pn4-prima}), which can be rewritten as
\begin{eqnarray} \nonumber
(P^4_n)^{\prime}&=&-\frac{1}{3}e^{-(3a+n)+b}(P^4_n)^2\left(1+3\gamma_0 \right)\\
                          &=&-2(1+3\gamma_0)P^4_n\left[\frac{1}{6}e^{-(3a+n)+b}P^4_n\right]
\end{eqnarray}

Using equation (\ref{a-prima}),
\begin{equation}
(P^4_n)^{\prime}=-2(1+3\gamma_0)P^4_n a^{\prime}
\end{equation}

This equation can be integrated easily, yielding
\begin{equation}\label{Pn4}
P^4_n=\alpha_1 e^{-2a(1+3\gamma_0)}
\end{equation}

where $\alpha_1$ is obtained by evaluating the previous expression at $y=0$: $\alpha_1=-k_5 e^{a_0(5+6\gamma_0)}\rho_0$. Now let us find a relation between the variables $n$ and $a$. Combining equations (\ref{a-prima}) and (\ref{n-prima}),
\begin{eqnarray} \nonumber
n^{\prime}&=&\frac{1}{6}e^{-(3a+n)+b}P^4_n\left[2+3\gamma_0 \right]\\ 
                &=&-a^{\prime}\left(2+3\gamma_0 \right)
\end{eqnarray}

If we integrate, we can be obtain an equation that directly relates $a$ with $n$, 
\begin{equation}
n=-(2+3\gamma_0)a+\alpha_2
\end{equation}

$\alpha_2$ is also obtained by evaluating $n$ and $a$ at $y=0$; assuming that the brane has an FRW metric, $N_0=1 \rightarrow n_0=0$, so that $\alpha_2=(2+3\gamma_0)a_0$. We can now substitute $P^4_n$ and $n$ into equation (\ref{a-prima})

\begin{eqnarray} \nonumber
a^{\prime}&=&\frac{1}{6}e^{-(3a+n)+b}P^4_n\\ \nonumber
&=&\frac{\alpha_1}{6} Exp\left[-(3a-(2+3\gamma_0)a+\alpha_2)+b-2a(1+3\gamma)_0\right] \\
&=&-\frac{1}{6}k_5 \rho_0 e^{3a_0(1+\gamma_0)}e^{-3a(1+\gamma_0)}e^b
\end{eqnarray}

Returning again to the variable $A=e^a$,
\begin{equation}
\frac{A^{\prime}}{A}=-\frac{1}{6}k_5 \rho_0 A_0^{3(1+\gamma_0)}  A^{-3(1+\gamma_0)}B
\end{equation}

Let us integrate this equation, considering that the dependence of the variable $B$ on $y$ is not significant in our equations, so we can consider only its dependence on $t$, that is, $B=B_0(t)$. The solution for the variable $A$ is
\begin{equation}
A=A_0\left[ 1-3(1+\gamma_0)\frac{1}{6}k_5\rho_0 B_0 y \right]^{\frac{1}{3(1+\gamma_0)}}
\end{equation}

The equation for $N$ is
\begin{eqnarray} \nonumber
N&=&e^{\alpha_2}A^{-(2+3\gamma_0)}\\
&=&\left[ 1-3(1+\gamma_0)\frac{1}{6}k_5\rho_0 B_0 y \right]^{\frac{-(2+3\gamma_0)}{3(1+\gamma_0)}}
\end{eqnarray}

Note that for $\gamma_0=-1$, the solution is the exponential, corresponding to the $AdS_5$, namely,
\begin{eqnarray}
	A&=&A_0 Exp\left[ -\frac{1}{6}k_5\rho_0 B_0 y \right]\\
	N&=&Exp\left[ -\frac{1}{6}k_5\rho_0 B_0 y \right]
\end{eqnarray}

In our model, we obtain power-law solutions for metric coefficients, as in \cite{CapillaValdepena2020,PerezGarciaUrena2014}, where we solved Einstein's equations directly. Note that the temporal evolution of our system is completely determined by the functions $A_0$, $B_0$, and $
\rho_0$. Note that the geometry of the 5-dimensional space is determined by the matter content on the brane, namely $\gamma_0$.

\section{ADM formalism in 5-d}

In this section we will work with the Hamiltonian formalism by means of foliating the 5-dimensional space into 3-dimensional surfaces moving through the 5-d space. For this purpose, we will define a two-dimensional space-time direction from which the induced metric is obtained; subsequently, the extrinsic curvature tensor is calculated and the action and Hamiltonian equations are writen .

\subsection{Induced metric in 5-D}

Just as the induced metric is defined in the ADM formalism, and following the notation of \cite{Jha2023}, we define the projection as 
\begin{equation}\label{induced}
h_{\mu\nu}=g_{\mu\nu}+n^{(0)}_{\mu}n^{(0)}_{\nu}-n^{(4)}_{\mu}n^{(4)}_{\nu},
\end{equation}

where $n^{(0)}_\mu$ and $n^{(4)}_\mu$ are the unit vectors in the $x^0$ and spatial $x^4$ directions. This means that we are performing a foliation in a direction that is a linear combination of the unit vectors, namely, $m^{\mu}=Nn^{(0)\mu}+Bn^{(4)\mu}$. We can write $h_{\mu\nu}$ more compactly by introducing a factor $\epsilon_{ij}$
\begin{equation}
h_{\mu\nu}=g_{\mu\nu}+\epsilon_{ij}n^{(i)}_{\mu}n^{(j)}_{\nu}
\end{equation}
Their corresponding contravariant and mixed tensors are, respectively,
\begin{eqnarray}
	h^{\mu\nu}&=&g^{\mu\nu}+\epsilon_{ij}n^{(i)\mu}n^{(j)\nu}\\ h^{\nu}_{\mu}&=&\delta^{\nu}_{\mu}+\epsilon_{ij}n^{(i)}_{\mu}n^{(j)\nu},
\end{eqnarray} 
where the indices $i,j$ take only the values $\{0,4\}$, whereas $\mu,\nu$ take the values $\{0, 1, 2, 3, 4\}$. In general,
\begin{center}
$\epsilon_{ij} = \begin{pmatrix}
	\epsilon_{00} & \epsilon_{04}  \\
	\epsilon_{40} & \epsilon_{44}
\end{pmatrix}$
\end{center}

Note that $n^{(0)}_\mu$ and $n^{(4)}_\mu$ are timelike and spacelike vectors, respectively; we also assume that they are normalized, so we have the first identity
\begin{equation}
n^{(i)}_\mu n^{(j)\mu}=-\epsilon^{ij}
\end{equation}

Multiplying this last equation by $\epsilon_{ik}$, and knowing that $\epsilon_{ik}\epsilon^{ij}=\delta^j_k$, we have
\begin{equation}
\epsilon_{ik} n^{(i)}_\mu n^{(j)\mu}=-\delta^j_k
\end{equation}
From the previous result, we see that the trace of $h$ is
\begin{equation}
	h^{\mu}_{\mu}=\delta^{\mu}_{\mu}+\epsilon_{ij}n^{(i)}_{\mu}n^{(j)\mu}=\delta^{\mu}_{\mu}-\delta^i_i=5-2=3
\end{equation}

These last two identities allow us to demonstrate the properties of the projector
\begin{eqnarray} \nonumber
	h_{\mu\nu}n^{(k)\nu}&=&\left[g_{\mu\nu}+\epsilon_{ij}n^{(i)}_{\mu}n^{(j)}_{\nu}\right]n^{(k)\nu}\\ \nonumber
		&=&g_{\mu\nu}n^{(k)\nu}+\epsilon_{ij}n^{(i)}_{\mu}n^{(j)}_{\nu}n^{(k)\nu}\\ \nonumber
		&=&n^{(k)}_{\mu}+n^{(i)}_{\mu}\left(\epsilon_{ij}n^{(j)}_{\nu}n^{(k)\nu}\right)\\ \nonumber
		&=&n^{(k)}_{\mu}+n^{(i)}_{\mu}\left(-\delta^k_i\right)\\
		&=&n^{(k)}_{\mu}-n^{(k)}_{\mu}=0
\end{eqnarray}
This indicates that $h_{\mu
u}$ defines a three-dimensional surface orthogonal to the vectors $n^{(k)
u}$. Now, let us verify another identity
\begin{eqnarray} \nonumber
	h_{\alpha\mu}h^{\alpha\nu}&=&\left[g_{\alpha\mu}+\epsilon_{ij}n^{(i)}_{\alpha}n^{(j)}_{\mu}\right]\left[g^{\alpha\nu}+\epsilon_{kl}n^{(k)\alpha}n^{(l)\nu} \right]\\ \nonumber
	&=&g^{\nu}_{\mu}+\epsilon_{kl}n^{(k)}_{\mu}n^{(l)\nu}+\epsilon_{ij}n^{(i)\nu}n^{(j)}_{\mu}+(\epsilon_{ij}n^{(i)}_{\alpha}n^{(k)\alpha})\epsilon_{kl}n^{(j)}_{\mu}n^{(l)\nu}\\ \nonumber
	&=&g^{\nu}_{\mu}+\epsilon_{jl}n^{(j)}_{\mu}n^{(l)\nu}+\epsilon_{ij}n^{(i)\nu}n^{(j)}_{\mu}+(-\delta^k_j)\epsilon_{kl}n^{(j)}_{\mu}n^{(l)\nu}\\
	&=&g^{\nu}_{\mu}+\epsilon_{ij}n^{(i)\nu}n^{(j)}_{\mu}=h^{\nu}_{\mu}
\end{eqnarray}
In the last step we used the symmetry property of $\epsilon$, namely, $\epsilon_{ij}=\epsilon_{ji}$

From these identities, we will construct the metric tensor. First, we define a vector in the temporal direction and the extra dimension as
\begin{equation}
	m^{\alpha}=\epsilon_{ij}N^{(i)}n^{(j)\alpha},
\end{equation}
where $N^{(0)}=N$ and $N^{(4)}=B$. Note that $\epsilon_{ij}$ lowers the index of the $N^{i}$, so that $N_{(0)}=N$ and $N_{(4)}=-B$. These definitions will be useful later.

\subsection{5-D Lagrangian}

Now, recalling the definition of the extrinsic curvature tensor as the projection of the covariant derivative of the unit vector $n^{(k)}_{\beta}$
\begin{equation}
K^{(k)}_{\mu\nu}\equiv -h^{\alpha}_{\mu}h^{\beta}_{\nu}\nabla_{\alpha}n^{(k)}_{\beta}
\end{equation}

These last definitions allow us to demonstrate the Gauss identities, so we begin by calculating the Riemann tensor on the 3-surface
\begin{equation}\label{3riemann}
\left[ D_\alpha, D_\beta \right]V^{\gamma} = \prescript{(3)}{}R^{\gamma}_{\mu\alpha\beta}V^{\mu}
\end{equation}

Let us first calculate $D_{\alpha}D_{\beta}V^{\gamma}$ and then interchange the indices. Let us use the properties of the projector
\begin{eqnarray} \nonumber
D_\alpha (D_\beta V^{\gamma})&=&h^{\nu}_{\beta}h^{\gamma}_{\rho}h^{\mu}_{\alpha}\nabla_\mu (D_\nu V^{\rho} ) \\ \nonumber
&=&h^{\mu}_{\alpha}h^{\nu}_{\beta}h^{\gamma}_{\rho}\nabla_\mu ( h^{\sigma}_{\nu}h^{\rho}_{\lambda}\nabla_\sigma V^{\lambda} ) \\
&=&h^{\mu}_{\alpha}h^{\nu}_{\beta}h^{\gamma}_{\rho}\nabla_\sigma (V^{\lambda})\nabla_\mu ( h^{\sigma}_{\nu}h^{\rho}_{\lambda} ) + h^{\mu}_{\alpha}h^{\sigma}_{\beta}h^{\gamma}_{\lambda}\nabla_\mu ( \nabla_\sigma V^{\lambda} )
\end{eqnarray}

Here we have used $h_{\alpha\mu}h^{\alpha\nu}=h^{\nu}_{\mu}$ to reduce the second term. Now, let us develop the first term on the right-hand side, taking into account that $h^{\nu}_{\mu}=\delta^{\nu}_{\mu}+\epsilon_{ij}n^{(i)}_{\mu}n^{(j)\nu}$, the identity $h^{\nu}_{\mu}n^{(k)\mu}=0$, and the definition of the curvature tensor $K^{(k)}_{\mu\nu}\equiv -h^{\alpha}_{\mu}h^{\beta}_{\nu}\nabla_{\alpha}n^{(k)}_{\beta}$
\begin{eqnarray} \nonumber
	h^{\mu}_{\alpha}h^{\nu}_{\beta}h^{\gamma}_{\rho}\nabla_\sigma (V^{\lambda})\nabla_\mu ( h^{\sigma}_{\nu}h^{\rho}_{\lambda}) )&=&h^{\mu}_{\alpha}h^{\sigma}_{\beta}h^{\gamma}_{\rho}\nabla_\sigma (V^{\lambda})\nabla_\mu ( h^{\rho}_{\lambda})+h^{\mu}_{\alpha}h^{\nu}_{\beta}h^{\gamma}_{\lambda}\nabla_\sigma (V^{\lambda})\nabla_\mu ( h^{\sigma}_{\nu})\\ \nonumber
	&=&\epsilon_{ij}n^{(i)}_{\lambda}h^{\mu}_{\alpha}h^{\sigma}_{\beta}h^{\gamma}_{\rho}\nabla_\sigma (V^{\lambda})\nabla_\mu (n^{(j)\rho})+\epsilon_{ij}n^{(j)\sigma}h^{\mu}_{\alpha}h^{\nu}_{\beta}h^{\gamma}_{\lambda}\nabla_\sigma(V^{\lambda})\nabla_\mu ( n^{(i)}_{\nu})\\
	&=&-\epsilon_{ij}K^{(j)\gamma}_{\alpha}h^{\sigma}_{\beta}n^{(i)}_{\lambda}\nabla_\sigma (V^{\lambda})+\epsilon_{ij}n^{(j)\sigma}h^{\gamma}_{\lambda}\nabla_\sigma(V^{\lambda})K^{(i)}_{\alpha\beta}
\end{eqnarray}

Note that the last term is symmetric in the indices $\alpha$ and $\beta$, so it will cancel when writing the commutator. Now, since $V^{\lambda}$ is a 3-dimensional vector, we have $n^{(i)}_\lambda V^{\lambda}=0$, so we can write $n^{(i)}_{\lambda}\nabla_\sigma(V^{\lambda})=-V^{\lambda}\nabla_\sigma(n^{(i)}_{\lambda})=- h^{\lambda}_{\delta}V^{\delta}\nabla_\sigma(n^{(i)}_{\lambda})$. Thus, we write
\begin{eqnarray} \nonumber
	h^{\mu}_{\alpha}h^{\nu}_{\beta}h^{\gamma}_{\rho}\nabla_\sigma (V^{\lambda})\nabla_\mu ( h^{\sigma}_{\nu}h^{\rho}_{\lambda}) )&=&\epsilon_{ij}K^{(j)\gamma}_{\alpha}h^{\sigma}_{\beta}h^{\lambda}_{\delta}V^{\delta}\nabla_\sigma (n^{(i)}_{\lambda}) \\
	&=&-\epsilon_{ij}K^{(j)\gamma}_{\alpha}K^{(i)}_{\beta \delta}V^{\delta}
\end{eqnarray}

Now, using convenient index changes, the commutator can be written as
\begin{eqnarray} \nonumber
	\left[ D_\alpha, D_\beta \right]V^{\gamma} &=& -\epsilon_{ij}\left[K^{(j)\gamma}_{\alpha}K^{(i)}_{\beta \delta}-K^{(j)\gamma}_{\beta}K^{(i)}_{\alpha \delta}\right]V^{\delta}+h^{\mu}_{\alpha}h^{\sigma}_{\beta}h^{\gamma}_{\lambda}\left[\nabla_\mu  \nabla_\sigma -\nabla_\sigma  \nabla_\mu\right] V^{\lambda}\\
	&=&-\epsilon_{ij}\left[K^{(j)\gamma}_{\alpha}K^{(i)}_{\beta \mu}-K^{(j)\gamma}_{\beta}K^{(i)}_{\alpha \mu}\right]V^{\mu}+h^{\rho}_{\alpha}h^{\sigma}_{\beta}h^{\gamma}_{\lambda}\prescript{(4)}{}R^{\lambda}_{\mu\rho\sigma} V^{\mu}
\end{eqnarray}

Comparing with equation (\ref{3riemann}), assuming that $V^{\mu}$ is an arbitrary vector and with suitable index manipulation, we arrive at the Gauss relation
\begin{eqnarray}
h^{\mu}_{\alpha}h^{\nu}_{\beta}h^{\gamma}_{\rho}h^{\sigma}_{\delta}\prescript{(4)}{}R^{\rho}_{\sigma\mu\nu}	&=&\prescript{(3)}{}R^{\gamma}_{\delta\alpha\beta}+\epsilon_{ij}\left[K^{(j)\gamma}_{\alpha}K^{(i)}_{\delta \beta}-K^{(j)\gamma}_{\beta}K^{(i)}_{\alpha \delta}\right]
\end{eqnarray}

The contracted versions follow directly using the properties of the induced metric, so we write the curvature scalar as
\begin{eqnarray}\label{gauss4}
	\prescript{(4)}{}R+2\epsilon_{ij}n^{(i)\mu}n^{(j)\nu}\prescript{(4)}{}R_{\mu\nu}&=&\prescript{(3)}{}R+\epsilon_{ij}\left[K^{(i)}K^{(j)}-K^{(j)\mu\nu}K^{(i)}_{\mu\nu}\right]
\end{eqnarray}

This is the scalar Gauss identity. Before writing the action, we need to work a little with the term $2\epsilon_{ij}n^{(i)\mu}n^{(j)\nu}\prescript{(3)}{}R_{\mu\nu}$. We can write this term using the commutator
\begin{eqnarray} \nonumber
	2\epsilon_{ij}n^{(i)\mu}n^{(j)\nu}\prescript{(3)}{}R_{\mu\nu}&=&2\epsilon_{ij}n^{(j)\nu}\left[ \nabla_\mu, \nabla_\nu \right]n^{(i)\mu} \\
	&=&2\epsilon_{ij}\left[ n^{(j)\nu}\nabla_\mu \nabla_\nu n^{(i)\mu} - n^{(j)\nu}\nabla_\nu \nabla_\mu n^{(i)\mu}  \right]
\end{eqnarray}

If we add and subtract the term $\nabla_\alpha  n^{(i)\mu}\nabla_\mu  n^{(j)\alpha}- \nabla_\alpha  n^{(i)\alpha}\nabla_\mu  n^{(j)\mu}$, with a little algebra, we obtain
\begin{eqnarray} \nonumber
	2\epsilon_{ij}n^{(i)\mu}n^{(j)\nu}\prescript{(3)}{}R_{\mu\nu}&=&2\epsilon_{ij}\left[ \nabla_\alpha(n^{(i)\mu}\nabla_\mu  n^{(j)\alpha} - n^{(i)\alpha}\nabla_\mu  n^{(j)\mu}) - \nabla_\alpha  n^{(i)\mu}\nabla_\mu  n^{(j)\alpha} + \nabla_\alpha  n^{(i)\alpha}\nabla_\mu  n^{(j)\mu}  \right]
\end{eqnarray}

The total divergence can be ignored in the action, and the remaining terms can be rewritten in terms of the extrinsic curvature tensor
\begin{eqnarray}
	2\epsilon_{ij}n^{(i)\mu}n^{(j)\nu}\prescript{(3)}{}R_{\mu\nu}&=&2\epsilon_{ij}\left[  K^{(i)}K^{(j)} - K^{(j)\mu\nu}K^{(i)}_{\mu\nu} \right]
\end{eqnarray}

Substituting into the Gauss identity (\ref{gauss4}),
\begin{eqnarray}
	\prescript{(4)}{}R &=&\prescript{(3)}{}R+\epsilon_{ij}\left[K^{(j)\mu\nu}K^{(i)}_{\mu\nu}-K^{(i)}K^{(j)}\right]
\end{eqnarray}

This is the main result of this section. We can now write the Einstein-Hilbert action with a cosmological constant component $\int^{t_2}_{t_1}dt \int_{\Sigma_t} d^3 x \sqrt{-g}\left(\prescript{(3)}{}R-2\Lambda_5\right)$
\begin{equation}
\int^{t_2}_{t_1}dt \int_{\Sigma_t} d^3 x \sqrt{-g}\left[ \prescript{(3)}{}R-2\Lambda_5+\epsilon_{ij}\left(K^{(j)\mu\nu}K^{(i)}_{\mu\nu}-K^{(i)}K^{(j)}\right) \right]
\end{equation}

The Lagrangian density is then written as
\begin{equation} \label{lag-dens}
	\mathcal{L} = \sqrt{-g}\left[ \prescript{(3)}{}R-2\Lambda_5+\epsilon_{ij}\left(K^{(j)\mu\nu}K^{(i)}_{\mu\nu}-K^{(i)}K^{(j)}\right) \right]
\end{equation}

This is a generalization of the Lagrangian in the ADM formalism. In the following section, we will find the explicit form of the curvature tensors in order to write the Hamiltonian.

\subsection{Contravariant vectors and Lie derivative}

The Lie derivative in a direction $m^{\mu}$ is defined as
\begin{equation}
	\mathcal{L}_m h_{\mu\nu}=m^{\alpha}\nabla_{\alpha}h_{\mu\nu}+h_{\alpha\nu}\nabla_{\mu}m^{\alpha}+h_{\mu\alpha}\nabla_{\nu}m^{\alpha}
\end{equation}

If we write an arbitrary vector in the $x^k$ direction, $m^{(k)\alpha}=N_{(k)}n^{(k)\alpha}$ (no index convention applies), with $k=\{0,4\}$
\begin{eqnarray}\nonumber
	\mathcal{L}^{(k)}_m h_{\mu\nu}&=&m^{(k)\alpha}\nabla_{\alpha}h_{\mu\nu}+h_{\alpha\nu}\nabla_{\mu}m^{(k)\alpha}+h_{\mu\alpha}\nabla_{\nu}m^{(k)\alpha} \\ \nonumber
	&=&N_{(k)}n^{(k)\alpha}\nabla_{\alpha}\left(\epsilon_{ij}n^{(i)}_\mu n^{(j)}_\nu \right)+h_{\alpha\nu}N_{(k)}\nabla_{\mu}n^{(k)\alpha}+h_{\mu\alpha}N_{(k)}\nabla_{\nu}n^{(k)\alpha} \\
	&&+h_{\alpha\nu}n^{(k)\alpha}\nabla_{\mu}N_{(k)}+h_{\mu\alpha}n^{(k)\alpha}\nabla_{\nu}N_{(k)}
\end{eqnarray}

The last two terms vanish due to the property $h^{\nu}_{\mu}n^{(k)\mu}=0$. If we multiply both sides of the equation by $h^{\mu}_{\delta}h^{\nu}_{\sigma}$ and assume that the Lie derivative is also a 3-dimensional tensor,
\begin{eqnarray}\nonumber
	\mathcal{L}^{(k)}_m h_{\delta\sigma}&=&h^{\mu}_{\delta}h^{\nu}_{\sigma}\mathcal{L}^{(k)}_m h_{\mu\nu} \\
	&=&h^{\mu}_{\delta}h^{\nu}_{\sigma}N_{(k)}n^{(k)\alpha}\nabla_{\alpha}\left(\epsilon_{ij}n^{(i)}_\mu n^{(j)}_\nu\right)+h^{\mu}_{\delta}h_{\alpha\sigma}N_{(k)}\nabla_{\mu}n^{(k)\alpha}+h^{\nu}_{\sigma}h_{\delta\alpha}N_{(k)}\nabla_{\nu}n^{(k)\alpha}
\end{eqnarray}

Again, the first term vanishes and the remaining two terms are written in terms of the extrinsic curvature tensor,
\begin{eqnarray}\nonumber
	\mathcal{L}^{(k)}_m h_{\delta\sigma}&=&N_{(k)}h_{\sigma\beta}\left(h^{\mu}_{\delta}h^{\beta}_{\alpha}N_{(k)}\nabla_{\mu}n^{(k)\alpha}\right)+N_{(k)}h_{\epsilon\delta}\left(h^{\nu}_{\sigma}h^{\epsilon}_{\alpha}N_{(k)}\nabla_{\nu}n^{(k)\alpha}\right) \\ \nonumber
	\mathcal{L}^{(k)}_m h_{\delta\sigma}&=&N_{(k)}h_{\sigma\beta}\left( -K^{(k)\beta}_{\delta} \right)+N_{(k)}h_{\epsilon\delta}\left( -K^{(k)\epsilon}_{\sigma}\right) \\
	\mathcal{L}^{(k)}_m h_{\delta\sigma}&=&-2N_{(k)} K^{(k)}_{\delta\sigma}
\end{eqnarray}

Recalling that the curvature tensor is symmetric in its indices, we finally arrive at
\begin{eqnarray} \label{Lie}
	\mathcal{L}^{(k)}_m h_{\mu\nu}&=&-2N_{(k)} K^{(k)}_{\mu\nu} \ \ \ \ \left[ \text{no sum in k}\right]
\end{eqnarray}

Now, let us focus on explicitly calculating the Lie derivative. For this purpose, we first need to define the unit vectors and the inverse matrix $g^{\mu\nu}$. Let us first perform the foliation in a direction that is a linear combination of the coordinates $x^0$ and $x^4$ and perpendicular to the hypersurface $\Sigma_t$, such that the line element is
\begin{eqnarray} \nonumber
ds^2&=&g_{\mu\nu}\left( N n^{(0)\mu}dx^0+B n^{(4)\mu}dx^4 \right) \left( N n^{(0)\nu}dx^0+B n^{(4)\nu}dx^4 \right) \\
&&+h_{ab}\left( N^a dx^0+B^a dx^4+dx^a \right) \left( N^b dx^0+B^b dx^4+dx^b \right)
\end{eqnarray}

Recalling the definition $\epsilon^{ij}=-n^{(i)}_{\mu}n^{(j)\mu}$
\begin{eqnarray} \nonumber
ds^2&=&\ \left(N^a N_a-\epsilon^{00}N^2\right)dx^0 dx^0 +\left( N_a B^a -\epsilon^{04}NB \right)dx^0 dx^4 \\ \nonumber
&& + \left(2N_a\right) dx^0 dx^a+h_{ab} dx^a dx^b + \left(2B_{a}\right) dx^4 dx^a   \\
&&  +\left(  B_a N^a -\epsilon^{40}BN \right)dx^4 dx^0+\left( B_a B^a-\epsilon^{44}B^2 \right) dx^4 dx^4
\end{eqnarray}

\begin{center}
	$g_{\mu \nu} = \begin{pmatrix}
		N^a N_{a}-\epsilon^{00}N^2 & \ \ \ \ \ \  N_a  \ \ \ \ \ \ &N_a B^a-\epsilon^{04}NB \\
		N_b & \ \ h_{ab}\ \  & B_b \\
		 B_a N^a-\epsilon^{40}BN & B_a & B_a B^a-\epsilon^{44}B^2
	\end{pmatrix}$
\end{center}

A direct verification allows us to confirm that the inverse of $g_{\mu \nu}$ is
\[
g^{\mu\nu} =
\begin{pmatrix}
	-\dfrac{\epsilon^{44}}{\epsilon N^2}
	&
	\dfrac{\epsilon^{44}}{\epsilon N^2}N^b
	-\dfrac{\epsilon^{04}}{\epsilon NB}B^b
	&
	\dfrac{\epsilon^{04}}{\epsilon NB}
	\\[4mm]
	
	\dfrac{\epsilon^{44}}{\epsilon N^2}N^a
	-\dfrac{\epsilon^{04}}{\epsilon NB}B^a
	&
	g^{ab}
	&
	-\dfrac{\epsilon^{04}}{\epsilon NB}N^a
	+\dfrac{\epsilon^{00}}{\epsilon B^2}B^a
	\\[4mm]
	
	\dfrac{\epsilon^{04}}{\epsilon NB}
	&
	-\dfrac{\epsilon^{04}}{\epsilon NB}N^b
	+\dfrac{\epsilon^{00}}{\epsilon B^2}B^b
	&
	-\dfrac{\epsilon^{00}}{\epsilon B^2}
\end{pmatrix}
\qquad
\]

where
\begin{eqnarray}
\epsilon&=&\epsilon^{00}\epsilon^{44}-(\epsilon^{04})^2 \\
g^{ab}&=&h^{ab}-\frac{\epsilon^{44}}{\epsilon N^2}N^aN^b+\frac{\epsilon^{04}}{\epsilon NB}\left(N^aB^b+B^aN^b\right)-\frac{\epsilon^{00}}{\epsilon B^2}B^aB^b	
\end{eqnarray}

Now, if we write the covariant vectors $n^{(0)}_\mu$ and $n^{(4)}_\mu$ as
\begin{eqnarray}
	n^{(0)}_\mu&=&(-N,0,0,0,0) \\
	n^{(4)}_\mu&=&(0,0,0,0,+B)
\end{eqnarray} 

we can then calculate $n^{(k)\mu}=g^{\mu\nu}n^{(k)}_\nu$
\begin{eqnarray}
	n^{(0)\mu}&=&\left[\dfrac{\epsilon^{44}}{\epsilon N},-\dfrac{\epsilon^{44}}{\epsilon N}N^a
	+\dfrac{\epsilon^{04}}{\epsilon B}B^a,-\dfrac{\epsilon^{04}}{\epsilon B}\right] \\
	\\
	n^{(4)\mu}&=&\left[ \dfrac{\epsilon^{04}}{\epsilon N}, -\dfrac{\epsilon^{04}}{\epsilon N}N^a
	+\dfrac{\epsilon^{00}}{\epsilon B}B^a, -\dfrac{\epsilon^{00}}{\epsilon B} \right]
\end{eqnarray} 

This choice necessarily requires $\epsilon^{44}/\epsilon=\epsilon^{00}$ and $\epsilon^{00}/\epsilon=\epsilon^{44}$, so, simplifying,
\begin{eqnarray}
	n^{(0)\mu}&=&\left[\dfrac{\epsilon^{00}}{N},-\dfrac{\epsilon^{00}N^a}{ N}
	+\dfrac{\epsilon^{04}}{\epsilon B}B^a,-\dfrac{\epsilon^{04}}{\epsilon B}\right] \\
	\\
	n^{(4)\mu}&=&\left[ \dfrac{\epsilon^{04}}{\epsilon N}, -\dfrac{\epsilon^{04}}{\epsilon N}N^a
	-\dfrac{\epsilon^{44}B^a}{B}, -\dfrac{\epsilon^{44}}{ B} \right]
\end{eqnarray}

Now we explicitly write $m^{(k)\mu}=N_{(k)}n^{(k)\mu}$ (there is no sum over $k$), so in each direction we have $m^{(0)\mu}=N_{(0)}n^{(0)\mu}$ and $m^{(4)\mu}=N_{(4)}n^{(4)\mu}$

Since our connection is torsion-free, we write the Lie derivative as
\begin{eqnarray} \nonumber
	\mathcal{L}^{(k)}_m h_{\mu\nu}&=&m^{(k)\alpha}D_{\alpha}h_{\mu\nu}+h_{\alpha\nu}D_{\mu}m^{(k)\alpha}+h_{\mu\alpha}D_{\nu}m^{(k)\alpha} \\
	&=&N_{(k)}\left[n^{(k)\alpha}D_{\alpha}h_{\mu\nu}+h_{\alpha\nu}D_{\mu}n^{(k)\alpha}+h_{\mu\alpha}D_{\nu}n^{(k)\alpha} \right]
\end{eqnarray}

We now compare with the expression found previously (\ref{Lie}), $\mathcal{L}^{(k)}_m h_{\mu\nu}=-2N_{(k)} K^{(k)}_{\mu\nu}$,
\begin{equation}
K^{(k)}_{ab}=-\frac{1}{2}\left[n^{(k)\alpha}D_{\alpha}h_{ab}+h_{c b}D_{a}n^{(k)c}+h_{a c}D_{b}n^{(k)c} \right]
\end{equation}

Note that the indices $\mu,\nu$ have been changed to $a,b$, because the extrinsic curvature tensor is an object on the surface $\Sigma_t$. The index $\alpha$ still takes the values $(0,1,2,3,4)$, but taking into account that $D_c h_{ab}=0$ on the 3-surface, we finally write the curvature tensor in the two directions $x^0$ and $x^4$ as
\begin{equation}
K^{(k)}_{ab}=-\frac{1}{2}\left[n^{(k)0}\partial_{0}h_{ab}+n^{(k)4}\partial_{4}h_{ab}+h_{c b}\partial_{a}n^{(k)c}+h_{a c}\partial_{b}n^{(k)c} \right]
\end{equation}

\subsection{Hamiltonian in a 5-D space}

Using the explicit form of the vectors $n^{(0)\mu}$ and $n^{(4)\mu}$, and analyzing the particular case $\epsilon^{00}=1$, $\epsilon^{44}=-1$ and $\epsilon^{04}=\epsilon^{40}=0$,
\begin{eqnarray}
	n^{(0)\mu}&=&\left[\dfrac{1}{N},-\dfrac{N^a}{ N}
	,0\right] \\
	\\
	n^{(4)\mu}&=&\left[0,
	-\dfrac{B^a}{B}, \dfrac{1}{ B} \right]
\end{eqnarray} 

The curvature tensors are
\begin{eqnarray}
K^{(0)}_{ab}&=&-\frac{1}{2N}\left(\partial_{0}h_{ab}-\partial_{a}N_b-\partial_{b}N_a \right) \\
K^{(4)}_{ab}&=&-\frac{1}{2B}\left(\partial_{4}h_{ab}-\partial_{a}B_b-\partial_{b}B_a \right)
\end{eqnarray}

These last equations are now more familiar from the ADM formalism. We are now in a position to write the Lagrangian density (\ref{lag-dens})
\begin{equation}
	\mathcal{L} = \sqrt{-g}\left[ \prescript{(3)}{}R-2\Lambda_5+K^{(0)ab}K^{(0)}_{ab}-K^{(0)}K^{(0)} -K^{(4)ab}K^{(4)}_{ab}+K^{(4)}K^{(4)}\right]
\end{equation}

It is time to define the canonical momenta, which are written as
\begin{equation}
	\Pi^{(0)ab}=\dfrac{\partial \mathcal{L}}{\partial (\partial_0 h_{ab})}, \ \ \    \Pi^{(4)ab}=\dfrac{\partial \mathcal{L}}{\partial (\partial_4 h_{ab})}
\end{equation}

Using the last two identities, and with a little algebra, we easily arrive at
\begin{eqnarray}
	\Pi^{(0)ab}&=&B\sqrt{h}\left(K^{(0)}h^{ab}-K^{(0)ab}\right) \Longrightarrow K^{(0)ab}=\dfrac{1}{2B\sqrt{h}}\left(\Pi^{(0)}h^{ab}-2\Pi^{(0)ab}\right)\\
	\Pi^{(4)ab}&=&N\sqrt{h}\left(-K^{(4)}h^{ab}+K^{(4)ab}\right) \Longrightarrow K^{(4)ab}=\dfrac{1}{2N\sqrt{h}}\left(2\Pi^{(4)ab}-\Pi^{(4)}h^{ab}\right)
\end{eqnarray}

We now write the Lagrangian density in terms of the canonical momenta
\begin{equation}\nonumber
	\mathcal{L} = BN\sqrt{h}\left(\prescript{(3)}{}R-2\Lambda_5\right)+\dfrac{N}{B\sqrt{h}}\left(\Pi^{(0)ab}\Pi^{(0)}_{ab}-\dfrac{1}{2}\Pi^{(0)}\Pi^{(0)}\right)-\dfrac{B}{N\sqrt{h}}\left(\Pi^{(4)ab}\Pi^{(4)}_{ab}-\dfrac{1}{2}\Pi^{(4)}\Pi^{(4)}\right)
\end{equation}

By definition, the Hamiltonian is $\mathcal{H}=\Pi^{(0)ab}\partial_0 h_{ab}+\Pi^{(4)ab}\partial_4 h_{ab}-\mathcal{L}$. It can be easily shown that the first two terms can be written as
\begin{eqnarray}
	\Pi^{(0)ab}\partial_0 h_{ab}&=&\dfrac{2N}{B\sqrt{h}}\left( \Pi^{(0)ab}\Pi^{(0)}_{ab}-\dfrac{1}{2}\Pi^{(0)}\Pi^{(0)} \right)+2\Pi^{(0)ab}\partial_a N_b\\
	\Pi^{(4)ab}\partial_4 h_{ab}&=&-\dfrac{2B}{N\sqrt{h}}\left(\Pi^{(4)ab}\Pi^{(4)}_{ab}-\dfrac{1}{2}\Pi^{(4)}\Pi^{(4)} \right)+2\Pi^{(4)ab}\partial_a B_b
\end{eqnarray}

Thus, the Hamiltonian density is
\begin{eqnarray} \nonumber
	\mathcal{H} &=& -BN\sqrt{h}\left(\prescript{(3)}{}R-2\Lambda_5\right)+\dfrac{N}{B\sqrt{h}}\left(\Pi^{(0)ab}\Pi^{(0)}_{ab}-\dfrac{1}{2}\Pi^{(0)}\Pi^{(0)}\right)-\dfrac{B}{N\sqrt{h}}\left(\Pi^{(4)ab}\Pi^{(4)}_{ab}-\dfrac{1}{2}\Pi^{(4)}\Pi^{(4)}\right) \\
	&&+2\Pi^{(0)ab}\partial_a N_b+2\Pi^{(4)ab}\partial_a B_b 
\end{eqnarray}

Let us write Hamilton's equations using the De Donder-Weyl formalism
\begin{eqnarray}
	\partial_0 h_{ab}&=&\dfrac{\partial \mathcal{H}}{\partial \Pi^{(0)ab}} = \frac{N}{B\sqrt{h}}\Pi^{(0)cd}\left( 2h_{ac}h_{bd}-h_{ab}h_{cd}\right)\\ 
	\partial_4 h_{ab}&=&\dfrac{\partial \mathcal{H}}{\partial \Pi^{(4)ab}}=-\frac{B}{N\sqrt{h}}\Pi^{(4)cd}\left( 2h_{ac}h_{bd}-h_{ab}h_{cd}\right) \\
	\partial_0 \Pi^{(0)ab}+\partial_4 \Pi^{(4)ab}&=&-\dfrac{\partial \mathcal{H}}{\partial h_{ab}}=-BN\sqrt{h}\left[ h^{ab}\left(\prescript{(3)}{}R-2\Lambda_5\right)+\frac{\partial}{\partial h_{ab}}\left(\prescript{(3)}{}R\right)\right]-\frac{\mathcal{H}}{2}h^{ab}
\end{eqnarray}

where, $\prescript{(3)}{}R$, in general, it does not have a reduced form as a function of $\kappa$. In the following section, we will examine a particular case where the 3-dimensional metric is symmetric and, in general, anisotropic.. We can note that in this configuration, for a symmetric induced metric, we have 18 equations, and if it is also diagonal, they are reduced to 9 equations. We will analyze this case in the following section.

%
%
%
%

\section{ADM formalism in a Randall-Sundrum configuration}

Consider a 3-dimensional Friedmann metric with curvature $\kappa$ such that $h_{ab}$ has a diagonal form: 
\begin{eqnarray}
h_{11}&=&\frac{A^2}{1-\kappa r^2}, \  \ h_{22}=A^2 r^2  \  \   \text{and} \ \  h_{33}=A^2 r^2 sin^2\theta \\
\prescript{(3)}{}R&=&\frac{6\kappa}{h_{11}(1-\kappa r^2)}
\end{eqnarray}

For this metric, is easy to show $h^{ab} \prescript{(3)}{}R+\frac{\partial}{\partial h_{ab}}\left(\prescript{(3)}{}R\right)=0$ The shift terms $N_a$, $B_a$ are zero. In this case, $h_{11}\Pi^{(k)11}=h_{22}\Pi^{(k)22}=h_{33}\Pi^{(k)33}$, so the Hamiltonian and its corresponding dynamical equations can be reduced to
\begin{eqnarray}
	\mathcal{H} &=& -BN\sqrt{h}\left(\prescript{(3)}{}R-2\Lambda_5\right)-\dfrac{3Nh^2_{11}}{2B\sqrt{h}}\left( \Pi^{(0)11}\right)^2+\dfrac{3Bh^2_{11}}{2N\sqrt{h}}\left( \Pi^{(4)11}\right)^2 \\
	\partial_0 h_{11}&=& -\frac{Nh^2_{11}}{B\sqrt{h}}\Pi^{(0)11} \\
	\partial_4 h_{11}&=&+\frac{Bh^2_{11}}{N\sqrt{h}}\Pi^{(4)11}  \\
	\partial_0 \Pi^{(0)11}+\partial_4 \Pi^{(4)11}&=&-2BN\Lambda_5\sqrt{h} h^{11}-\frac{\mathcal{H}}{2}h^{11}
\end{eqnarray}

These equations are the same as in the previous section, but with the difference that  $N^{\prime}$ and $\dot{B}$ are not present; for that reason, the Hamiltonian function is not zero here. These equations hold only on the hypersurface, so the Friedmann equation must be obtained directly from the Hamiltonian.
\begin{eqnarray}
	\frac{h^2_{11}}{6BN\sqrt{h}}\mathcal{H}&=&-\left[\frac{\partial_0 h_{11}}{2N}\right]^2 +\left[\frac{\partial_4 h_{11}}{2B}\right]^2-\frac{h_{11}\kappa}{1-\kappa r^2}+\frac{1}{3}\Lambda_5 h^2_{11} \nonumber
\end{eqnarray}

If we change $h_{11}=\frac{A^2}{1-\kappa r^2}$,
\begin{eqnarray}
	\frac{A^2}{6BN\sqrt{h}}\mathcal{H}&=&-\left[\frac{A\dot{A}}{N}\right]^2 +\left[\frac{A A^{\prime}}{B}\right]^2-A^2\kappa+\frac{1}{3}\Lambda_5 A^4 \nonumber
\end{eqnarray}

This last equation is the Friedmann equation, as in previous sections, but only for a very specific form of $\mathcal{H}$. The Friedmann equation cannot be obtained directly in this formalism, because there are no equations equivalent to the $G_{00}$, $G_{44}$ and $G_{04}$. 

%

\section{Conclusions}\label{section5}

In this work, we studied the Hamiltonian formalism in a five-dimensional brane model. By performing the $1+4$ splitting of spacetime, we obtain the evolution equations in time and along the extra coordinate, defining a Hamiltonian within the De Donder--Weyl formalism. The equations are solved under a particular ansatz, and the explicit form of the scale factors $A$ and $N$ as functions of $y$ is obtained, which does not follow directly from the Einstein equations. This Hamiltonian formalism has the drawback that the equation equivalent to $G_{04}$ does not appear, and therefore the Hamiltonian constraint $\mathcal{H}$ must be added by hand in order to recover the Friedmann equation. We also considered the ADM formalism, with the difference that the hypersurface $\Sigma_t$ is embedded in the five-dimensional bulk, and the foliation is performed along a direction that is a linear combination of the vectors $n^{(0)\mu}$ and $n^{(4)\mu}$. This system reduces the four Einstein equations to only.

\section*{ACKNOWLEDGMENTS}

 We are grateful to the \textit{Secretaría de Ciencias Humanidades, Tecnología e Innovación} (SECIHTI) for the financial support provided through the postdoctoral scholarship with CVU number 218255. Many calculations where done by Symbolic Program REDUCE 3.8.

\end{document}